\documentclass[a4paper,11pt]{scrartcl}
\usepackage[inner=2.5cm,outer=2.5cm,top=1.9cm,bottom=1.9cm,includeheadfoot]{geometry}
\usepackage[utf8]{inputenc}
\usepackage[T1]{fontenc}
\usepackage{graphicx}
\usepackage[detect-family,separate-uncertainty]{siunitx}
\usepackage{hyperref}
\makeatletter
\g@addto@macro{\UrlBreaks}{\do\/\do\-\do\.}
\makeatother

\usepackage[UKenglish]{isodate}
\cleanlookdateon
\usepackage{booktabs}
\usepackage{amssymb}

\usepackage[cmex10]{amsmath}
\usepackage{authblk}
\title{On temporal correlations in high--resolution frequency counting}

\date{}

\author[1]{Tim Dunker\thanks{tdu \{--at--\} justervesenet.no}}
\author[1,2]{Harald Hauglin\thanks{hha \{--at--\} justervesenet.no}}
\author[3]{Ole Petter R\o nningen\thanks{ole.petter \{--at--\} ictec.com}}
\footnotesize
\affil[1]{Justervesenet, P.O. Box 170, 2027 Kjeller, Norway}
\affil[2]{UNIK --- University Graduate Center, P.O. Box 70, 2027 Kjeller, Norway}
\affil[3]{ICTEC AS, Drammensveien 127, 0277 Oslo, Norway}

\usepackage[natbib=true,bibstyle=numeric,citestyle=numeric,backend=bibtex,maxcitenames=2,sortcites,maxbibnames=1000]{biblatex}
\usepackage[babel,english=british]{csquotes}
\DeclareFieldFormat[article]{title}{#1} 

\defbibheading{bibliography}{\section*{References}}

\renewbibmacro*{journal+issuetitle}{%
  \usebibmacro{journal}%
  \setunit*{\addspace}%
  \iffieldundef{series}
    {}
    {\newunit
     \printfield{series}%
     \setunit{\addspace}}%
  \printfield{volume}%
  \iffieldundef{number}
     {}
      {\mkbibparens{\printfield{number}}}%
   \setunit{\addcomma\space}%
   \printfield{eid}%
  \setunit{\addspace}%
  \usebibmacro{issue+date}%
  \setunit{\addcolon\space}%
  \usebibmacro{issue}%
\iffieldundef{addendum}{}
\setunit{\addcomma}
  \newunit}

\renewbibmacro{in:}{%
  \ifentrytype{article}{}{\printtext{\bibstring{in}\intitlepunct}}}

\DeclareFieldFormat{doi}{%
  \mkbibacro{DOI}\addcolon\space
    \ifhyperref
      {\href{http://dx.doi.org/#1}{\nolinkurl{#1}}}
      {\nolinkurl{#1}}}

\DeclareFieldFormat{url}{%
  \mkbibacro{URL}\addcolon\space
    \ifhyperref
      {\href{#1}{\nolinkurl{#1}}}%
      {\nolinkurl{#1}}}

\DeclareBibliographyCategory{badbreaks}
\addtocategory{badbreaks}{ctan}

\AtEveryBibitem{%
  \ifcategory{badbreaks}
    {\defcounter{biburllcpenalty}{9000}}
    {}}

\begin{document}

\maketitle

\begin{abstract}
We analyze noise properties of time series of frequency data from different counting modes of a Keysight 53230A frequency counter. We use a 10\,MHz reference signal from a passive hydrogen maser connected via phase--stable Huber+Suhner Sucoflex 104 cables to the reference and input connectors of the counter. We find that the high resolution gap--free (``CONT'') frequency counting process imposes long--term correlations in the output data, resulting in a modified Allan deviation MDEV$\sim\tau^{-1/2}$-characteristic of random walk phase noise. Equally important, the CONT mode results in a frequency bias. In contrast, the counter's undocumented raw continuous mode (``RCON'') yields unbiased frequency stability estimates with white phase noise characteristics, MDEV$\sim\tau^{-3/2}$, and of a magnitude consistent with the counter’s 20\,ps single--shot resolution.
Furthermore, we demonstrate that a 100--point running average filter in conjunction with the RCON mode yields resolution enhanced frequency estimates with flicker phase noise characteristics, MDEV$\sim\tau^{-1}$. For instance, the counter's built--in moving--average function can be used. The improved noise characteristics of the averaged RCON mode versus the CONT mode imply that the former mode yields frequency estimates with improved confidence for a given measurement time.
\end{abstract}

\section{Introduction}
\citet{Rubiola2005} and \citet{Dawkinsetal2007} described frequency counting methods and averaging processes, as well as their uncertainties. We investigate the CONT and RCON measurement modes of a widely used frequency counter---the Keysight 53230A. We measure a \SI{10}{\mega\hertz} reference signal against itself with this counter, and analyze the resulting frequency offset and autocorrelation function. We compare the different measurement modes to each other and determine which mode provides the most reliable frequency estimate. Furthermore, we measure the same signal with a Pendulum CNT--91 frequency counter and compare the results to those obtained with the Keysight counter's CONT and RCON modes.

The CONT measurement mode ``configures the counter for continuous, resolution--enhanced, gap--free
measurements`` \citep[p. 71]{Keysight}. According to Keysight, the CONT mode is required to achieve the best accuracy with respect to the Allan deviation calculated from frequency measurements \citep[p. 206]{Keysight}. However, we do not know exactly how the CONT mode operates. The RCON measurement mode is an undocumented, raw continuous frequency measurement mode.

The Keysight 53230A is a $\Pi$-counter if it operates in RCON mode. If it operates in RCON mode with internal averaging, the counter is a $\Lambda$-counter \citep{Benkleretal2015}. Different frequency counters use different techniques to reduce the influence of phase noise on frequency estimates \citep{Benkleretal2015}. One such method is internal averaging over a given time interval. We therefore look at different sampling intervals, and also apply a moving average to see how averaging affects the frequency estimates and the temporal correlations of the frequency measurements.

Temporal correlations in frequency measurements can be induced by a measurement mode, and may not be representative of the device under test. In particular, the autocorrelation function \citep[e.g.]{ShumwayStoffer2011} is a suitable tool to analyze the dependence between consecutive frequency measurements.

\section{Experimental setup}
\begin{figure}[!t]
\centering
\includegraphics[width=0.9\textwidth]{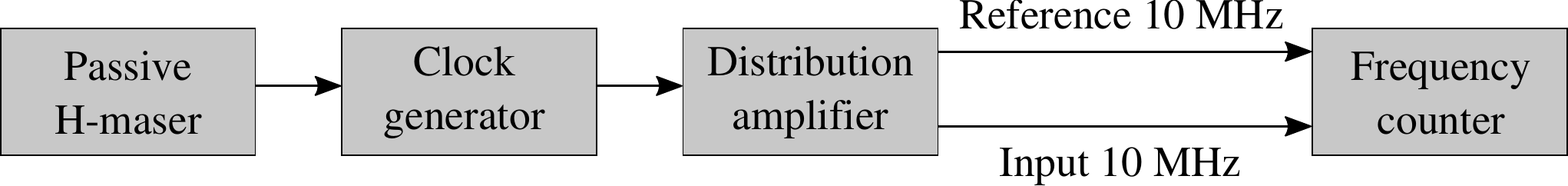}
\caption{Experimental setup}
\label{fig:setup}
\end{figure}

Figure \ref{fig:setup} is a sketch of our experimental setup. We connected a \SI{10}{\mega\hertz} signal from a passive hydrogen maser (Vremya-Ch VCH--1008) to a TimeTech 10274 distribution amplifier via SpectraDynamics HROG--10 clock generator. We coupled this \SI{10}{\mega\hertz} signal with phase-stable Huber+Suhner Sucoflex 104 cables to the reference and input connectors of a Keysight 53230A frequency counter \citep{Keysight}. 

We measured the frequency with the Keysight 53230A counter in CONT mode at time resolutions of \SI{10}{\second} and \SI{0.1}{\second}, and in the undocumented RCON mode with a time resolution of \SI{0.1}{\second}. We also computed a 100--point moving average of the RCON data measured with a time resolution of \SI{0.1}{\second}. We used the internal storage option provided by the Keysight 53230A counter, except for the CONT \SI{10}{\second} data, which we acquired remotely with TimeLab \citep{TimeLab}. When data are acquired with TimeLab, the default driver of the Keysight 53230A counter in TimeLab sends a ``READ?'' SCPI command to the counter, which triggers a new measurement, resulting in frequency measurements that are not gap--free. Instead, the ``R?'' SCPI command should be used, which does not trigger a new measurement. We could not detect any systematic difference between data acquired with TimeLab and those stored internally in the counter.

For comparison, we repeated the measurement at a sampling interval of \SI{0.1}{\second} with a Pendulum CNT--91 frequency counter in frequency back--to--back (FBTB) mode  \citep{Pendulum}.

\section{Results and discussion}

\subsection{Frequency estimates}
\begin{figure}[!t]
\centering
\includegraphics[width=0.9\textwidth]{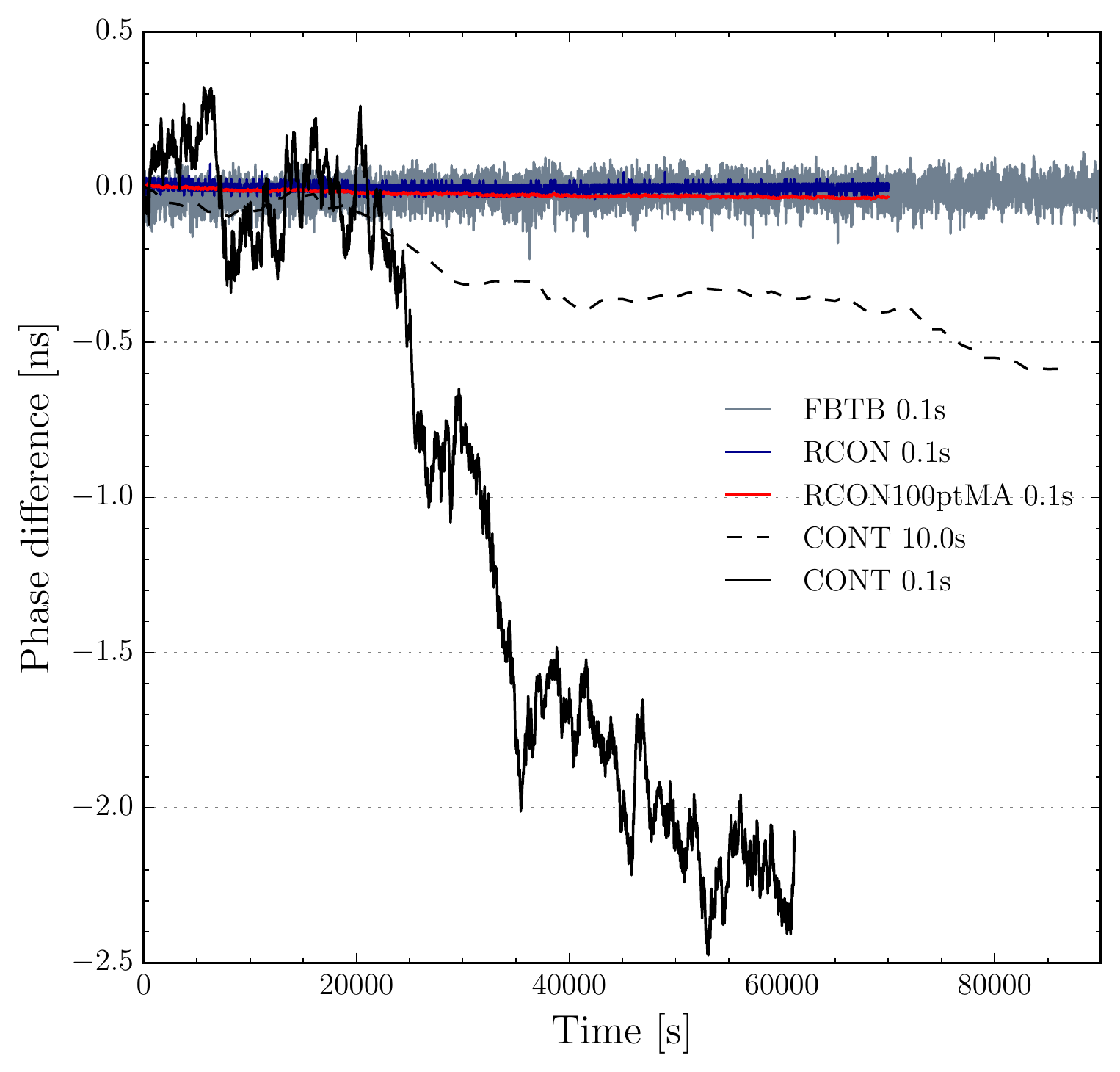}
\caption{Phase difference measured with the same experimental setup (Fig. \ref{fig:setup}) and different counters and modes: CONT \SI{0.1}{\second} (black, solid), CONT \SI{10.0}{\second} (black, dashed), RCON \SI{0.1}{\second} (blue), RCON \SI{0.1}{\second} with a 100--point moving average (red), and CNT--91 in frequency back--to--back mode and a temporal resolution of \SI{0.1}{\second} (grey). Only every 100th value is shown here. The CONT \SI{10.0}{\second} data were acquired with TimeLab.}
\label{fig:phase}
\end{figure}

Because we measure the reference signal at \SI{10}{\mega\hertz} against itself, the measured relative frequency offset should be \num{0}  in any measurement mode and at any sampling interval for a sufficiently long measurement period. From frequency measurements in different modes of a Keysight 53230A counter, we computed the phase difference. The results are shown in Fig. \ref{fig:phase}. For comparison, we also show a similar measurement made with a Pendulum CNT--91 counter in frequency back--to--back mode. As we expect from the specifications \citep{Keysight}, the Keysight counter exhibits less noise than the Pendulum CNT--91 \citep{Pendulum}. 

\begin{table}[!t]
   \centering
   \caption{Relative frequency offset, $\triangle f / f$, computed from frequency measurements in five different measurement modes. Compare with phase differences shown in Fig. \ref{fig:phase}.}
   \label{tab:freqoff}
   \begin{tabular}{lll}
   \toprule
    Instrument &  Measurement mode & $\triangle f / f$ \\
    \midrule
    Keysight 53230A & CONT \SI{10.0}{\second} & \num{-6.1e-15} \\
    Keysight 53230A & CONT \SI{0.1}{\second} & \num{-5.0e-14} \\
    Keysight 53230A & RCON \SI{0.1}{\second} & \num{-1.0e-16} \\
    Keysight 53230A & RCON100ptMA \SI{0.1}{\second} & \num{-4.9e-16} \\
    Pendulum CNT--91 & FBTB \SI{0.1}{\second} & \num{1.6e-16} \\
    \bottomrule
   \end{tabular}
\end{table}

We have fitted a least--squares linear regression to the phase difference data to obtain the relative frequency offset. We summarized these values in Table \ref{tab:freqoff}. Surprisingly, the CONT mode measurements yield a frequency bias at both sampling intervals. These are much larger than the relative frequency offsets measured in RCON mode or with the CNT--91 frequency counter. It is important to note that we cannot ascertain that the relative frequency offsets measured in RCON, RCON100ptMA, and FBTB mode are significantly different from \num{0}. 

The results from the frequency offsets (Table \ref{tab:freqoff}) show that the RCON mode yields a better frequency estimate than the CONT mode, even if we increase the sampling interval in CONT mode from \SI{0.1}{\second} to \SI{10}{\second}. 

If we increase the sampling interval from \SI{0.1}{\second} to \SI{10.0}{\second} in CONT mode, we obtain a smaller relative frequency offset, but is still different from 0. In particular, the RCON mode yields better results, with and without a moving--average filter.

\subsection{Autocorrelation function}
\begin{figure}[!t]
\centering
\includegraphics[width=0.9\textwidth]{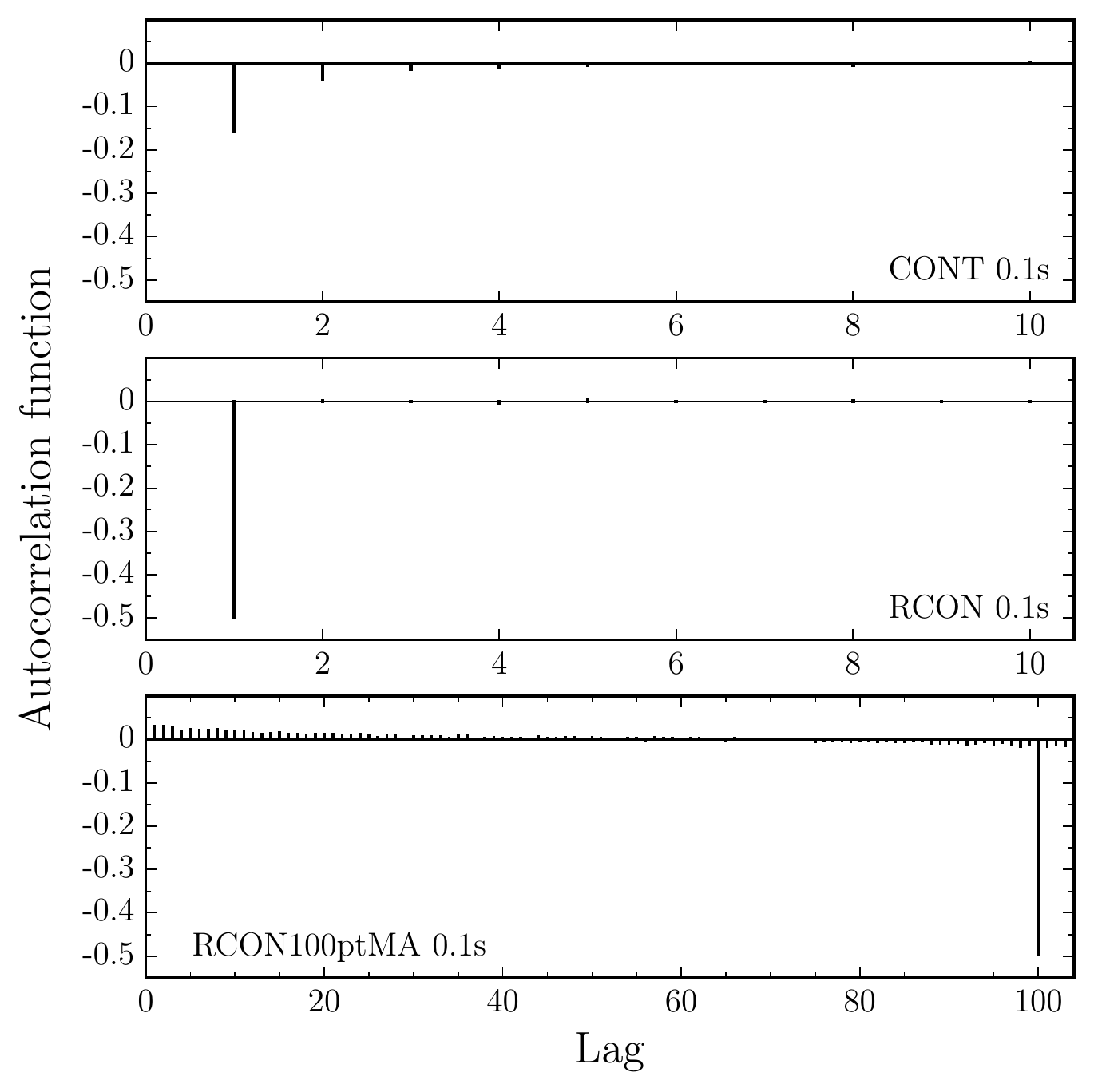}
\caption{Autocorrelation function for the measurement in different modes, made with a Keysight 53230A frequency counter. Top panel: CONT mode with \SI{0.1}{\second} time resolution. Middle panel: RCON mode with \SI{0.1}{\second} time resolution. Bottom panel: RCON mode with a time resolution of \SI{0.1}{\second} and a 100--point moving average. Note the different scales for the lag.}
\label{fig:autocorr}
\end{figure}

\citet[pp. 45 to 46]{Riley2008} pointed out that the dominant noise process in a time series of fractional frequency data can be identified using the lag \num{1} autocorrelation, $\rho_1$. If the data have not been differenced, white phase noise yields $\rho_1 = -1/2$, flicker frequency noise yields $\rho_1 = 1/3$, and white frequency noise yields $\rho_1 = 0$ \citep{Riley2008}. 

An autocorrelation of $\rho_1 = -1/2$ at lag \num{1} can be interpreted as there being a \SI{50}{\percent} chance of measuring a value smaller than average at $t=1$ if one measured a larger-than-average value at $t=0$.

For the RCON \SI{0.1}{\second} and CNT--91 FBTB \SI{0.1}{\second} measurements, the autocorrelation at lag 1 is $\rho_1 = -1/2$ (not shown). Thus, the dominating process is white phase noise. This is what we expect. The autocorrelation functions of the CONT \SI{10.0}{\second}, CONT \SI{0.1}{\second}, and RCON \SI{0.1}{\second} (with 100--point moving average) time series are shown in Fig. \ref{fig:autocorr}. The behaviour of the CONT mode is surprising. We see that the CONT mode results in temporal correlations at lag 1, which clearly are not due to white phase noise. In addition, the autocorrelation does not vanish for lags 2 and 3. Judging from the values of $\rho_1$, the dominant noise process seems to be a combination of flicker phase noise and random--walk phase noise.

When we apply a 100--point moving average to the RCON data with \SI{0.1}{\second} time resolution, the characteristics of the autocorrelation function do not change (Fig. \ref{fig:autocorr}, bottom panel). We now get an autocorrelation of $\rho_{100} = -1/2$. A finite averaging window does not lead to long temporal correlations that we see in CONT mode.

\subsection{Allan deviation}
\begin{figure}[!t]
\centering
\includegraphics[width=0.9\textwidth]{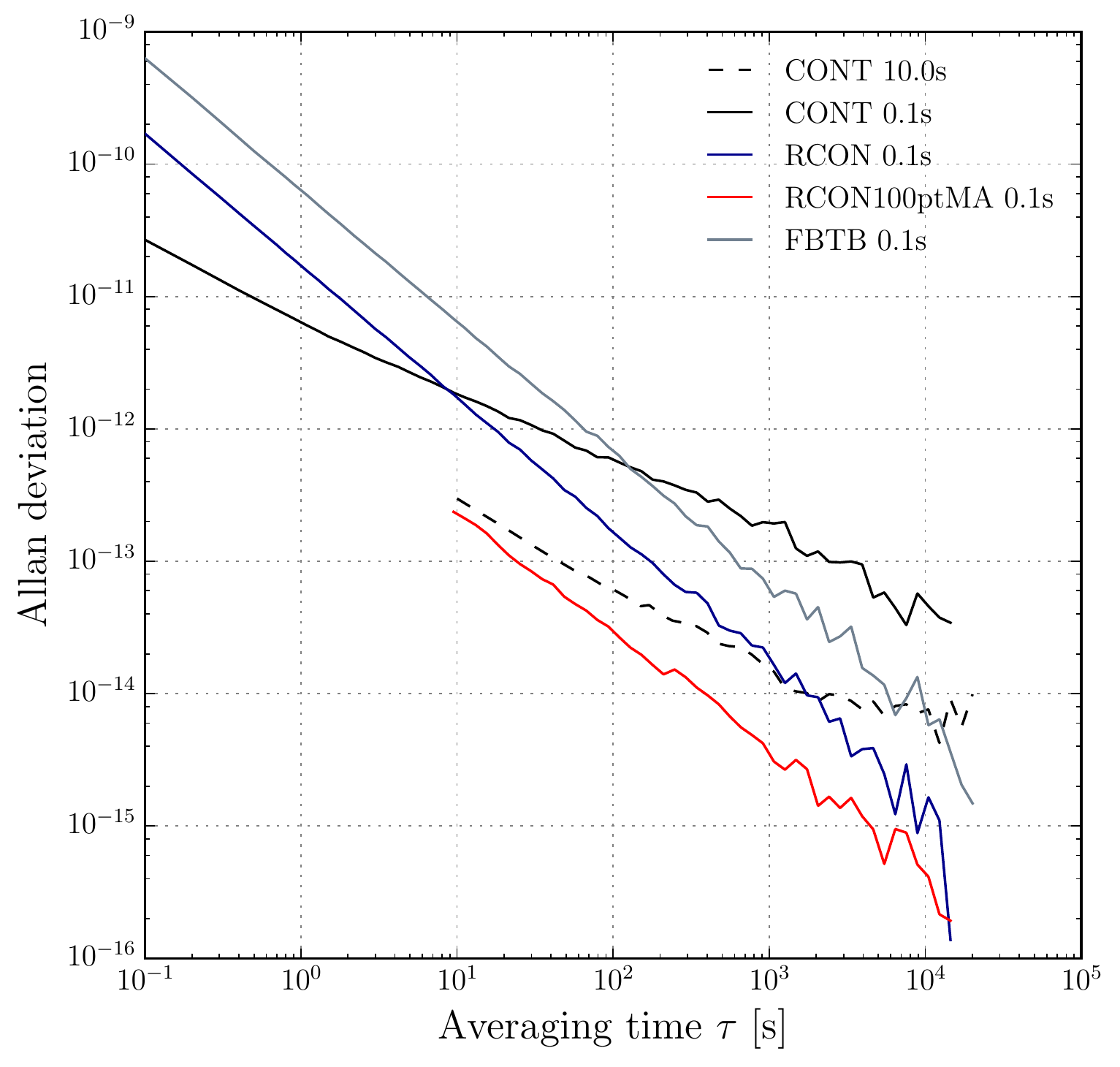}
\caption{Allan deviation of the time series shown in Fig. \ref{fig:phase}: CONT \SI{0.1}{\second} (black, solid), CONT \SI{10.0}{\second} (black, dashed), RCON \SI{0.1}{\second} (blue), RCON \SI{0.1}{\second} with a 100--point moving average (red), and CNT--91 in frequency back--to--back mode and a temporal resolution of \SI{0.1}{\second} (grey).}
\label{fig:adev}
\end{figure}

\begin{figure}[!t]
\centering
\includegraphics[width=0.9\textwidth]{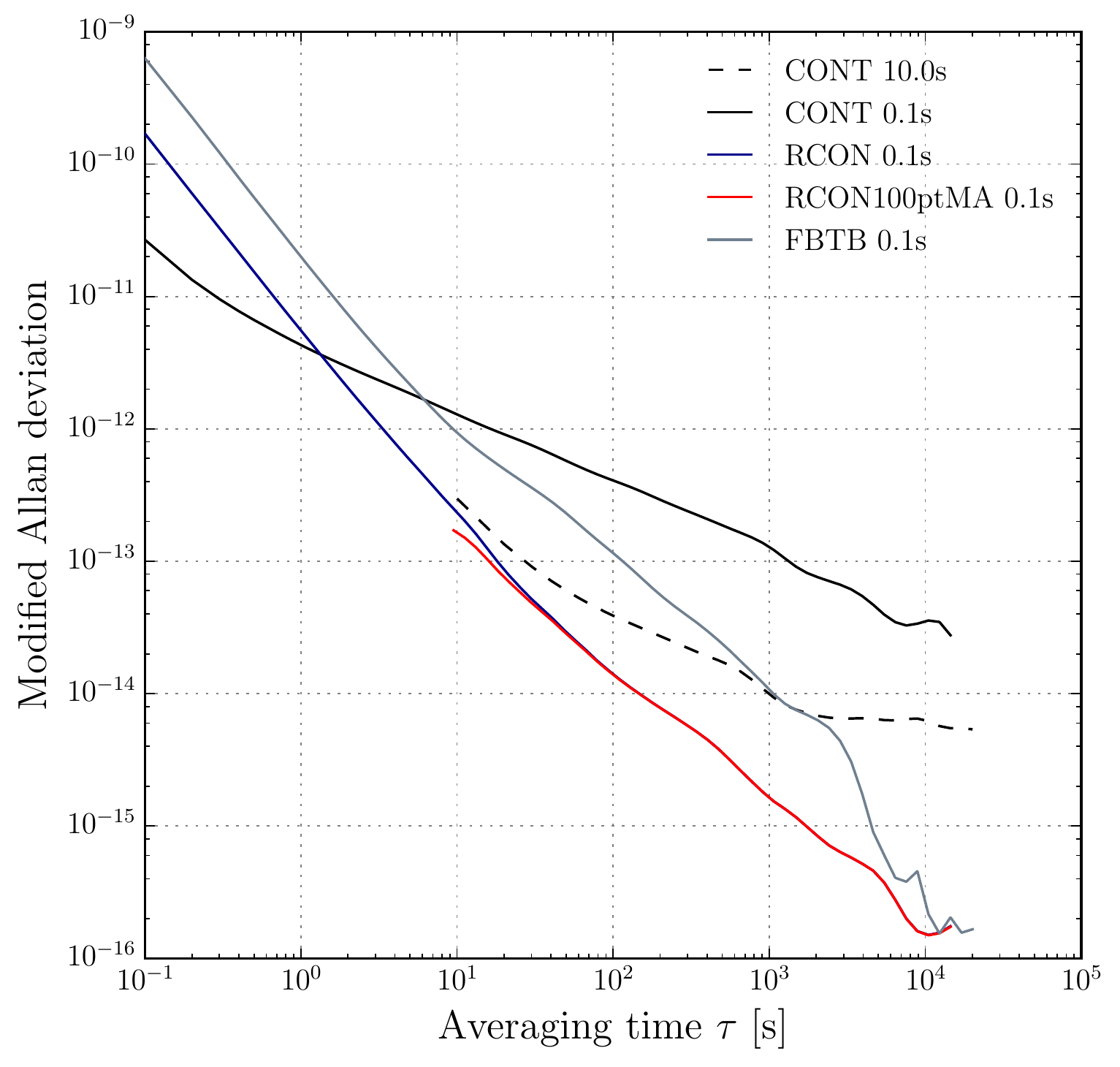}
\caption{Modified Allan deviation of the time series shown in Fig. \ref{fig:phase}: CONT \SI{0.1}{\second} (black, solid), CONT \SI{10.0}{\second} (black, dashed), RCON \SI{0.1}{\second} (blue), RCON \SI{0.1}{\second} with a 100--point moving average (red), and CNT--91 in frequency back--to--back mode and a temporal resolution of \SI{0.1}{\second} (grey).}
\label{fig:mdev}
\end{figure}

Figure \ref{fig:adev} shows the Allan deviation of the frequency measurements, and Fig. \ref{fig:mdev} shows the modified Allan deviation. From Fig. \ref{fig:adev}, we see that the RCON 100--point moving average time series with \SI{0.1}{\second} temporal resolution is approximately one order of magnitude better than RCON \SI{0.1}{\second} without moving average. This is consistent with the central limit theorem. We also see that, for averaging times $\tau \gtrsim \SI{10}{\second}$, the RCON mode provides a frequency estimate with better stability than the CONT mode.

The CONT mode frequency measurements result in a modified Allan deviation with an approximate slope of $\tau^{-1/2}$, which is characteristic of random--walk phase noise. We confirmed this result by calculating the first differences of the CONT \SI{0.1}{\second} time series and computing its autocorrelation (not shown). We obtained $\rho_1(d=1) = -1/2$, which corresponds to random--walk phase noise \citep{Riley2008}.

The undocumented RCON mode exhibits white phase noise characteristics, as does the Pendulum CNT--91 in frequency back--to--back mode. The modified Allan deviation of these measurements follows a slope of $\tau^{-3/2}$.

When we apply a 100--point moving average to the RCON \SI{0.1}{\second}--measurements, the modified Allan deviation follows a $\tau^{-1}$ slope. That is, we obtain resolution--enhanced frequency estimates with flicker phase noise characteristics. This is also true if we use the Keysight counter's built--in function for a 100--point moving average (not shown).

\section{Conclusions}
To characterize a Keysight 53230A frequency counter, we measured a \SI{10}{\mega\hertz} reference signal against itself, using the counter's different measurement modes (CONT and RCON). We compared these measurements to frequency back--to--back measurements made with a Pendulum CNT--91 counter.

We find that the Keysight 53230A's CONT mode, which is a ``resolution--enhanced, gap--free'' \citep{Keysight} frequency measurement mode, yields a frequency bias (error), even for measurement periods of roughly one day. The CONT mode exhibits random--walk phase noise, and the autocorrelation at lag 1 is not $-1/2$, as is typical for white phase noise. The autocorrelation for lag 2 does not vanish, leading to a temporal correlation in the frequency measurements. The frequency bias is present also for longer sampling intervals (\SI{10.0}{\second} instead of \SI{0.1}{\second}).

In contrast, the undocumented RCON mode does not show any frequency bias, and shows no temporal correlation at lag $\geq 2$. The modified Allan deviation exhibits white phase noise characteristics. The results are consistent with measurements made with a Pendulum CNT--91 frequency counter. 

Using the undocumented RCON mode, we can obtain frequency estimates with better stability and without bias. The resolution can be enhanced using the counter's built--in moving--average function.

%

%

\section*{Acknowledgments}
We used the Python package ``allantools'' by Anders Wallin, Danny Price, Cantwell G. Carson, and Fr\'ed\'eric Meynadier. We also used TimeLab by John Miles, whom we thank for his helpful advice.

\printbibliography

\end{document}